\documentclass[journal]{new-aiaa}
\usepackage[utf8]{inputenc}

\usepackage{graphicx}
\usepackage{amsmath}
\usepackage[version=4]{mhchem}
\usepackage{siunitx}
\usepackage{longtable,tabularx}
\usepackage{float}
\usepackage{subcaption}
\usepackage{multirow}

\usepackage[absolute,overlay,showboxes]{textpos}

\usepackage{overpic}

\usepackage{xcolor}					
\usepackage{soul}
\newcommand{\red}[1]{\textcolor{black}{#1}}
\newcommand{\blue}[1]{\textcolor{black}{#1}}
\newcommand{\green}[1]{\textcolor{black}{#1}}

\title{On the application of incomplete FWH surfaces \\ for aeroacoustic predictions}

\author{Tulio R. Ricciardi \footnote{Research Assistant, Department of Energy} and William R. Wolf\footnote{Assistant Professor, Department of Energy}}
\affil{University of Campinas, Campinas, SP, Brazil, 13083-860}
\author{Philippe Spalart \footnote{Boeing Senior Technical Fellow (retired)}}
\affil{Boeing Commercial Airplanes, Seattle, WA, 98124, USA}

\begin{document}

\maketitle

\section{Motivation}
\label{sec:intro}

The numerical prediction of aeroacoustic noise from complex airframe configurations, such as landing gears, high-lift devices and cavities, poses two major challenges which include the accurate evaluation of noise sources and their subsequent acoustic propagation to observer positions. The former is achieved through high-fidelity unsteady CFD calculations while the latter is typically carried out by using an acoustic analogy. The Ffowcs Williams and Hawkings (FWH) equation \cite{FWH1969} is commonly solved for aeroacoustic predictions, although several issues with respect to its application are still debated in the literature.  
One discussion concerns the role of quadrupole sources in airframe noise problems and their direct influence on the noise prediction \cite{WolfJFM2012, Spalart2013, Wolf2015}. To account for these sources, it is necessary to perform expensive flowfield volume integrations or to employ a permeable FWH surface surrounding the flow region where quadrupoles are non-negligible.
For low Mach number flows, typical of airframe noise problems, the common practice is to neglect the calculation of volume quadrupole sources since their noise radiation is usually less significant than that of surface dipole and monopole sources \cite{Curle1955}. A recent discussion on the accuracy of solid and permeable surfaces for airframe noise predictions is provided by Spalart et al. \cite{Spalart2019}.

In practical applications where the quadrupoles may be relevant, they are already accounted for by the use of permeable FWH surfaces. These integration surfaces should be placed as close as possible to the major sources (airfoils, landing gears, jets) to avoid dissipation by the numerical methods and mesh coarsening. At the same time, they should be placed as far as possible to avoid contamination by spurious nonlinear hydrodynamic sources crossing the faces as well as to avoid a hard truncation of volumetric sources \cite{Mitchell1999}. One naive approach to overcome the contamination issue is to simply ignore the downstream surface, employing a partially open (incomplete) integration surface \cite{Mendez2013}. On the other hand, multiple closing surfaces (the endcap approach) can be employed with a solution averaging, in terms of complex values, to filter uncorrelated hydrodynamic sources. In this case, the more coherent acoustic waves should be less affected by the phase averaging and the noise prediction is generally improved \cite{Shur2005,Mendez2013,Mahesh2014,Wright2015}. Other approaches evaluate additional surface terms to avoid the spurious noise generation in order to achieve better acoustic predictions \cite{Lockard2005,Rahier2015,ZhouWangWang2021}. Despite multiple efforts, a robust endcap strategy and other correction approaches still depend strongly on empiricism.

\blue{The motivation for the current study regards incomplete FWH surfaces, where a point of concern has been raised during the analysis of a realistic nose landing gear where only the fuselage half bottom was simulated in order to save computational resources \cite{Ricciardi2021}. In this case, the solid FWH surface was not entirely closed due to geometrical constraints. However, the derivation of the FWH equation and its proper solution rely on the application of closed surfaces through use of the Heaviside function. Since several studies in the aeroacoustics literature inappropriately employ open FWH surfaces, either solid or permeable, in the current work we present non-exhaustive results for two different model problems to highlight in what conditions an incomplete permeable surface is still a valid approach. By incomplete surface, here, we refer to a finite patch in a theoretically infinite plane as will be shown in the results section. This surface setup can be useful for airframe noise predictions of high-lift devices, landing gears and cavities, where observers of interest are typically in flyover or sideline positions as part of the certification of an airliner. For these problems, a single finite planar surface may provide accurate noise predictions while simplifying the permeable FWH surface setup. In this sense, the mesh for a complex CFD calculation can be designed with adaptive refinement applied only at a particular flow region, for example, below the slat in a multi-element airfoil, or under the landing gear in a full aircraft configuration. Moreover, the present approach avoids contamination from the back surface where quadrupole noise sources may cross the integration surface leading to spurious noise generation due to hard truncation. Hence, the planar surface setup does not require any special treatment of quadrupole sources or permeable surface corrections.}

\section{Methodology}

For a uniform freestream flow, the frequency domain FWH acoustic analogy \cite{Lockard2000,Lockard2002} can be written as the following boundary integral equation
\begin{equation}
\hat{p}(\mathbf{x}^{o},\omega) \, H(f) = - \int_{f=0} \left[ \mbox{i} \omega \hat{Q}(\mathbf{x}^{s},\omega)G_{c}(\mathbf{x}^{o},\mathbf{x}^{s}) + \hat{F}_{i}(\mathbf{x}^{s},\omega)\frac{\partial G_{c}(\mathbf{x}^{o},\mathbf{x}^{s})}{\partial x^{s}_{i}} \right] dS
- \int_{f>0} \hat{T}_{ij}(\mathbf{x}^{s},\omega) \frac{\partial^{2} G_{c}(\mathbf{x}^{o},\mathbf{x}^{s})}{\partial x^{s}_{i}\partial x^{s}_{j}} dV \mbox{,}
\label{eq:integral_FWH}
\end{equation}
where the pressure fluctuations $\hat{p}$ can be computed for a frequency $\omega$ at a particular observer position $\mathbf{x}^{o}=(x_1^{o},x_2^{o},x_3^{o})=(x^{o},y^{o},z^{o})$. The noise sources are computed at $\mathbf{x}^{s}=(x_1^{s},x_2^{s},x_3^{s})=(x^{s},y^{s},z^{s})$ and hats $\hat{\bullet}$ denote Fourier transformed quantities. The right hand side integrals are evaluated at the FWH surface, $f=0$, and along the flow region external to the FWH surface, $f>0$. The aeroacoustic predictions rely on the calculation of the quadrupole, dipole and monopole noise sources, which are given by
\begin{equation}
\blue{
T_{ij} = ( \rho u_{i} u_{j} + P_{ij} - c_{0}^{2} \rho' \delta_{ij} ) \mbox{ ,}}
\label{eq:quadrupole}
\end{equation}
\begin{equation}
\blue{F_{i} = ( P_{ij} + \rho(u_{i} - 2U_{i})u_{j} + \rho_{0}U_{i}U_{j} )\, n_{j}} \,\,\,\mbox{ , and}
\label{eq:dipole}
\end{equation}
\begin{equation}
\blue{Q = ( \rho u_{i} - \rho_{0}U_{i} )\, n_{i}} \mbox{ ,}
\label{eq:monopole}
\end{equation}
respectively. In the noise sources above, $u_i$ is the fluid velocity vector, $\rho$ is the density, with primes indicating fluctuations, $P_{ij}$ represents the normal and shear stresses, including the pressure and viscous stresses, $\rho_{0}$ and $c_0$ are the freestream density and speed of sound, respectively, and $\delta_{ij}$ is the Kronecker delta. For high Reynolds number flows, the effects of viscous stresses can be neglected and $P_{ij}= p \delta_{ij}$. The term $n_i$ represents the $i$-th component of the outward surface normal and $U_i$ is the uniform freestream velocity vector.

In the problems analyzed in this work, the freestream velocity is considered aligned with the $x$-Cartesian coordinate. Hence, the freestream Mach number is defined as $M = U_1/c_0$ and
flow effects on the acoustic propagation are accounted for by the following convective Green's functions:
\begin{equation}
G_{c}(\mathbf{x}^{o},\mathbf{x}^{s}) = \frac{\mbox{i}}{4\beta} \; e^{\mbox{i} M \frac{k}{\beta^2} \bar{x}} \; H_{0}^{(2)} \left( \frac{k}{\beta^2} R_{2D} \right) \; \,\,\, \mbox{, for 2D acoustic radiation, and}
\end{equation}
\begin{equation}
G_{c}(\mathbf{x}^{o},\mathbf{x}^{s}) = \frac{1}{4 \pi R_{3D}} \; e^{-\mbox{i} \frac{k}{\beta^2} \left[R_{3D} - M \bar{x} \right]} \mbox{ , for 3D acoustic radiation.}
\end{equation}
In these equations, the 2D and 3D source-observer distances are given by $R_{2D} = \sqrt{(x^{o}-x^{s})^{2} + \beta^{2} \left[(y^{o}-y^{s})^{2}\right]}$ and $R_{3D} = \sqrt{(x^{o}-x^{s})^{2} + \beta^{2} \left[(y^{o}-y^{s})^{2}+(z^{o}-z^{s})^{2}\right]}$, respectively. The Prandtl-Glauert correction term is $\beta = \sqrt{1-M^2}$,  $\bar{x} =  x^{o} - x^{s}$, and the acoustic wavenumber is defined as $k=\omega/c_0$. For 2D problems, $H_0^{(2)}$ represents the Hankel function of the second kind and order 0. 

\section{Results}

In this section, noise prediction results are presented for two different model problems, including a monopole source immersed in a uniform flow and the two-dimensional flow over a NACA0012 airfoil at low Reynolds number. The focus of the present analysis is on determining the conditions for which an incomplete permeable FWH surface provides accurate acoustic predictions in the frequency domain.

\subsection{Monopole source immersed in uniform flow}

The first problem consists of the 3D noise generation by a monopole immersed in a uniform flow \cite{Lockard2002}. The acoustic potential is defined as 
\begin{equation}
\phi(\mathbf{x}^{o},t) = A \frac{1}{4 \pi R_{3D}} e^{\mbox{i}[\omega t - \frac{k}{\beta}(R_{3D} - M \bar{x})]} \mbox{ .}
\end{equation}
Through calculation of the potential, the acoustic pressure, velocity and density are evaluated as $p'=-\rho_0 (\partial \phi/\partial t + U_1 \partial \phi/\partial x_1)$, $u'_i = \partial \phi/\partial x_i$ and $\rho' = p'/c_0^2$, respectively. Then, the noise sources are computed along a permeable FWH surface and, subsequently, are Fourier transformed.
By this example, it is possible to discuss the effects of incomplete surfaces in the FWH analogy. 
Typically, the frequency domain FWH equation is solved by means of a boundary integral formulation, such as Eq. \ref{eq:integral_FWH}, on a closed surface to evaluate the acoustic pressure at any observer position in the farfield. If the observers are placed inside the closed surface, their pressure fluctuations should be null due to the application of the Heaviside function, providing a consistency check to the analysis. \blue{A similar check can be done with the time domain FWH equation. However, this previous formulation requires the calculation of source terms at retarded times, being more cumbersome than the frequency domain approach, as discussed by Lockard \cite{Lockard2002}.}

\begin{figure}[H]
	\centering
	\begin{subfigure}{0.99\textwidth}
		\includegraphics[width=0.99\textwidth]{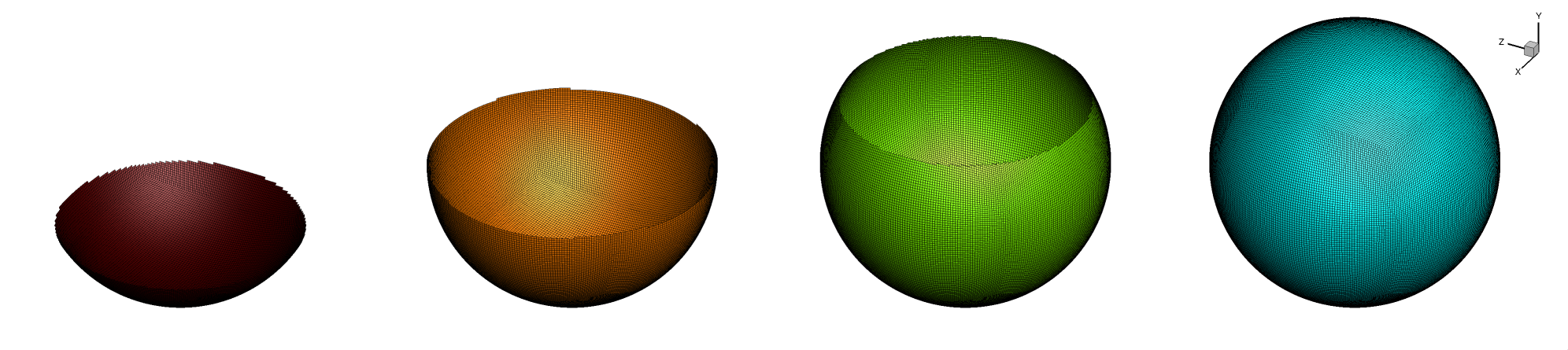}
		\subcaption{Different open (incomplete) and closed spheres employed as permeable FWH surfaces.}
	\end{subfigure}
	\begin{subfigure}{0.48\textwidth}
		\includegraphics[width=1.0\textwidth]{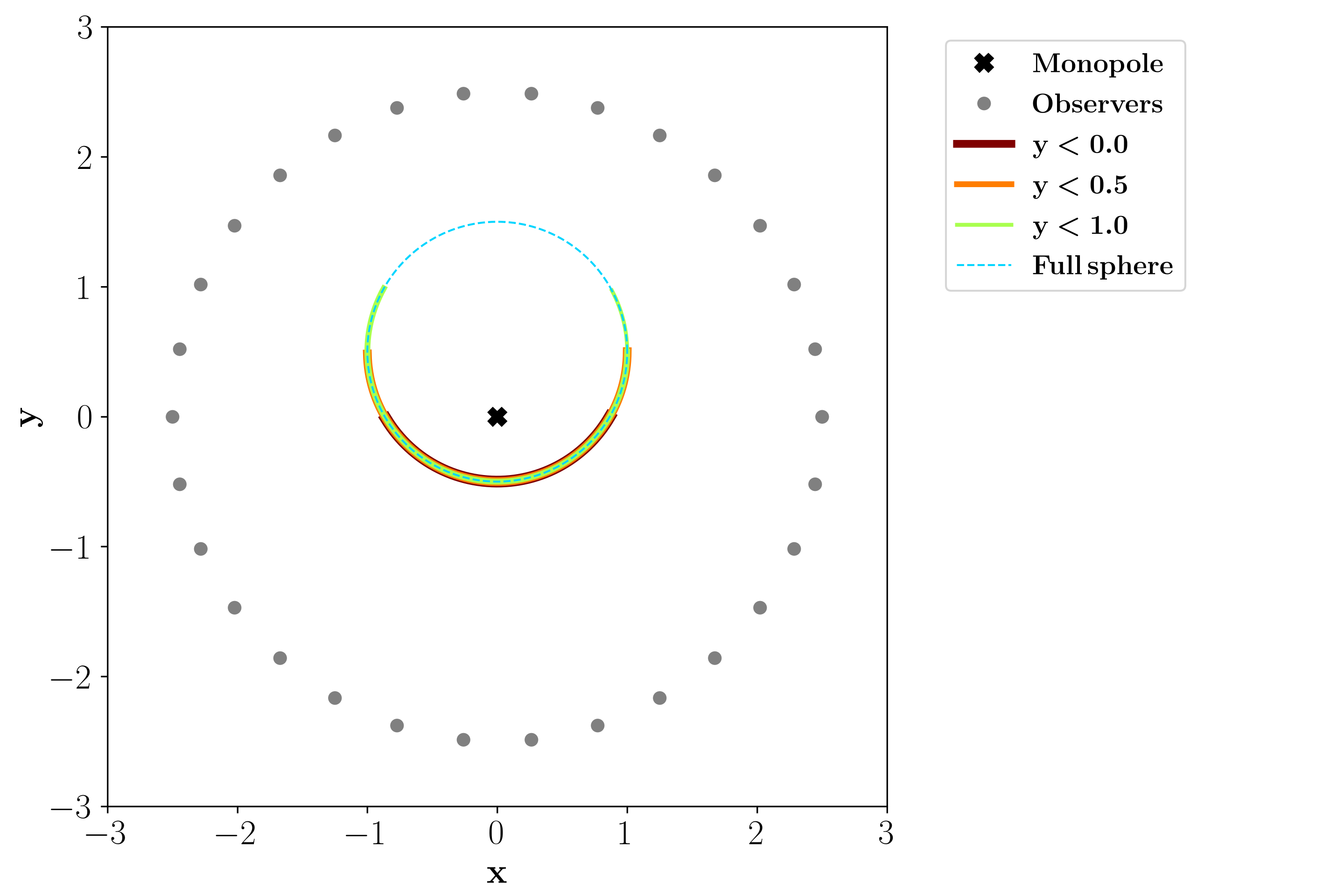}
	\subcaption{Permeable FWH surfaces, source and observer positions.}
	\end{subfigure}
	\begin{subfigure}{0.48\textwidth}
	\includegraphics[width=1.0\textwidth]{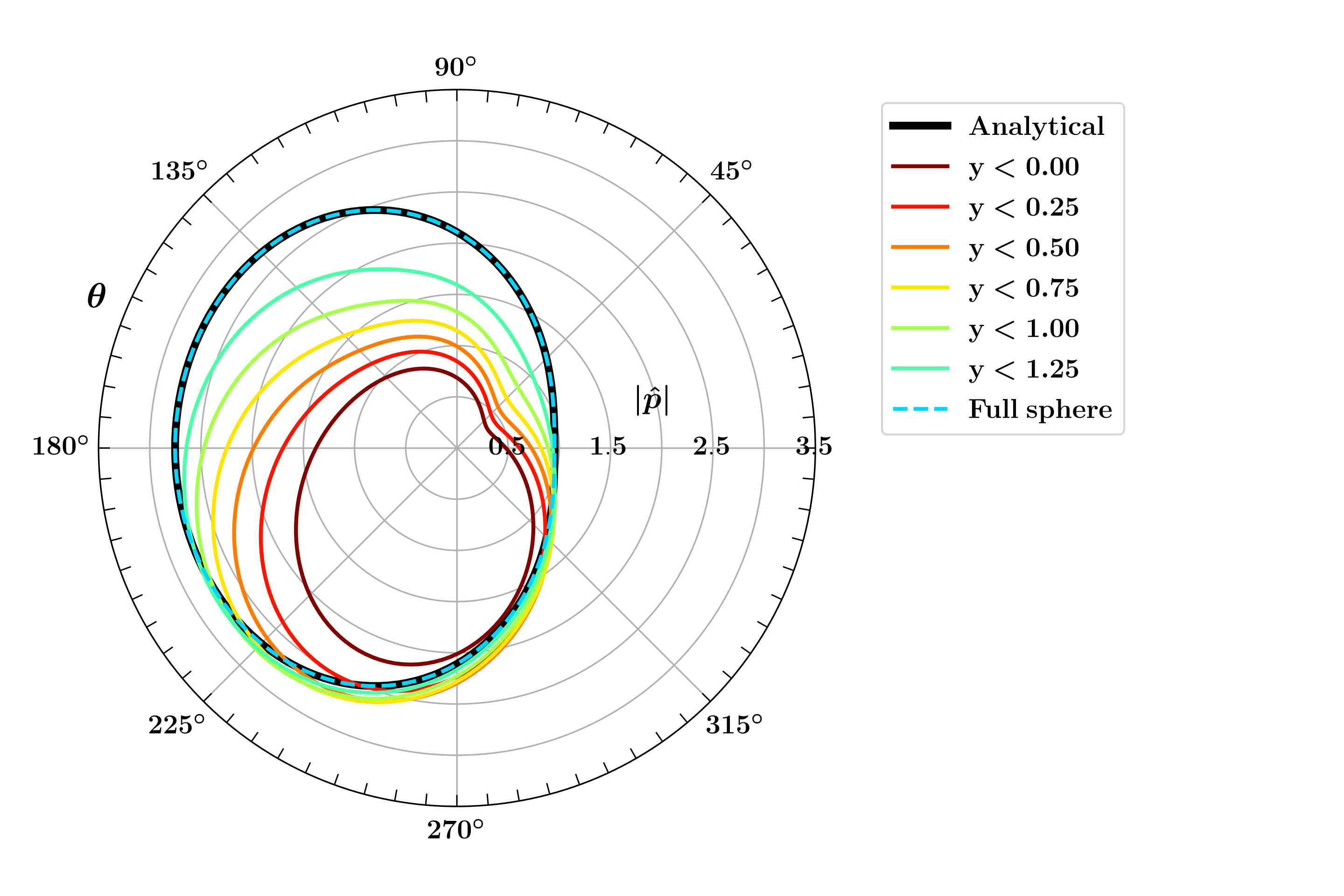}
	\subcaption{Directivity plots for different permeable FWH surfaces.}
	\end{subfigure}
	\caption{Analysis of closed and open spherical permeable FWH surfaces for monopole source immersed in $\boldsymbol{M=0.5}$ uniform flow.}
	\label{fig:truncated_monopole}
\end{figure}

To illustrate the issue previously discussed, a monopole with $k = 1$ immersed in a uniform flow with Mach number $M = 0.5$ is placed at the origin of the 3D Cartesian system. Then, the acoustic pressure is evaluated at a circle of observers with distance $r_o = 2.5$ measured from the origin, in the $z = 0.0$ plane. The noise prediction employs a spherical permeable surface with radius $r_s = 1.0$, centered on $x = 0.0$ and $y = +0.5$. The sphere is then truncated at different values of constant $y$ planes, as illustrated in Figs. \ref{fig:truncated_monopole}(a) and (b). This means that closed and open permeable surfaces (complete and incomplete spheres) are employed for the noise radiation as shown in the figure. Results of the noise predictions are presented in Fig. \ref{fig:truncated_monopole}(c). The closed permeable FWH surface shows excellent agreement with the analytical solution. For the incomplete spheres, the solutions show that, despite the permeable surfaces being placed between the monopole source and some observers, the noise is not accurately computed. Nonetheless, it is much better at flyover positions, around $\theta \approx 270$ deg, where the surfaces intersect the line-of-sight between source and observers. As more elements are added to the permeable surfaces, numerical results converge to the analytical solution. In this case, it is shown that even the FWH surface elements placed above the monopole source, at positive $y$ locations, contribute to the noise radiated to flyover observers at negative $y$ positions. \blue{This occurs because each surface element is composed by equivalent monopole and dipole sources as shown in Eq. \ref{eq:integral_FWH}. Therefore, the individual elements that compose the FWH surface radiate with directivities of the elementary sources and it is their constructive/destructive interference that provides the total noise emission at a specific observer location.} It is clear that solution convergence is faster for flyover observers, which are on the opposite side where the surface is truncated, indicating that directional effects are still important.

\begin{figure}[H]
	\centering
	\begin{subfigure}{0.45\textwidth}
		\includegraphics[width=0.99\textwidth]{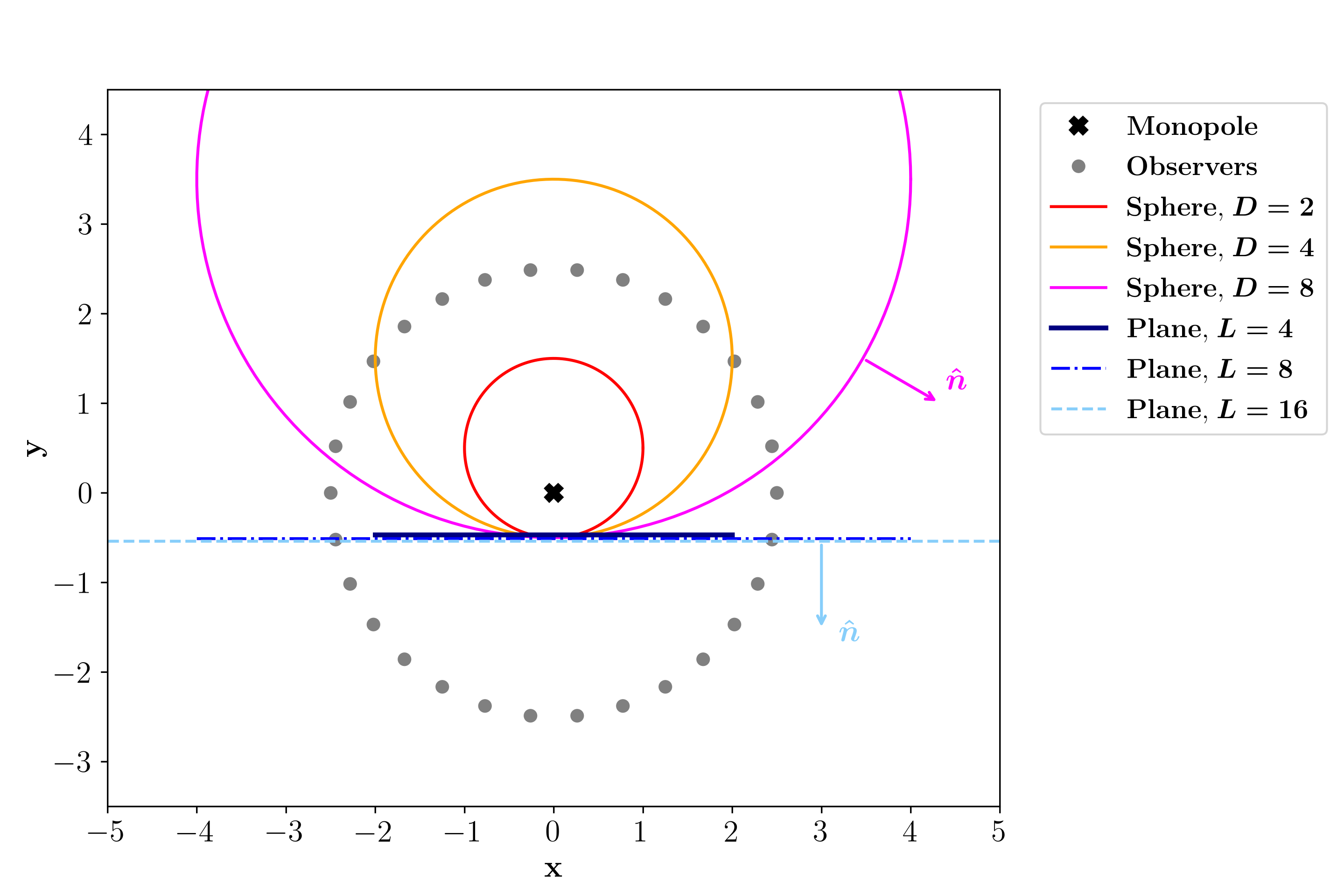}
		\subcaption{Closed and open FWH surfaces analyzed.}
	\end{subfigure}
	\begin{subfigure}{0.45\textwidth}
		\includegraphics[width=0.99\textwidth]{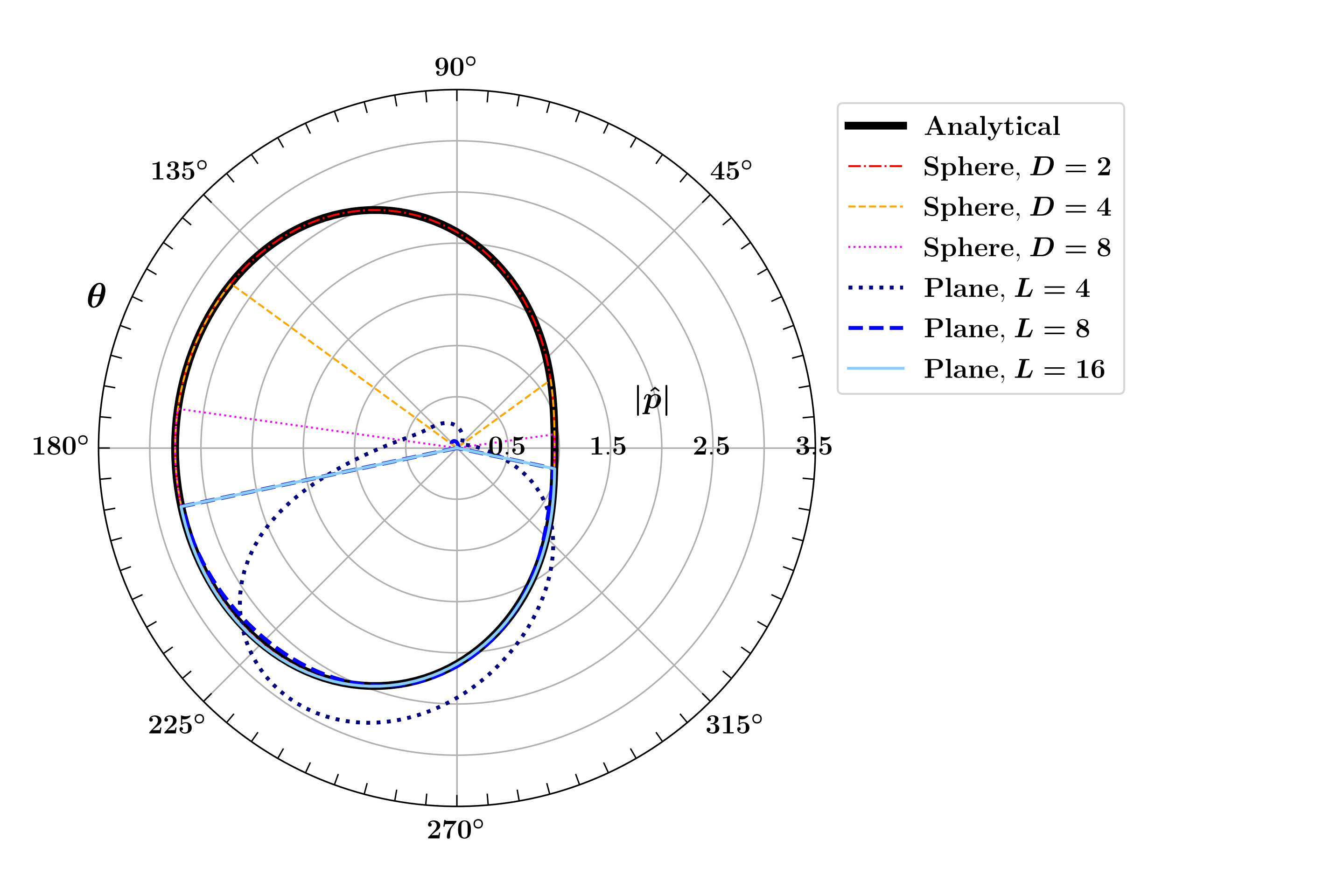}
		\subcaption{Directivities for closed and open surfaces.}
	\end{subfigure}
	\caption{Analysis of closed (spherical) and incomplete (planar) permeable FWH surfaces for monopole source immersed in $\boldsymbol{M = 0.5}$ uniform flow.}
	\label{fig:plane_monopole}
\end{figure}

For the present monopole problem, the magnitudes of source terms computed on the entire spherical permeable surface are similar, with some directional effects in the $x$ direction due to the mean flow effects and some small differences due to the offset of the permeable surface with respect to the $y$ axis. However, in realistic problems, the source magnitudes may decay significantly for some surface elements depending on numerical dissipation in the CFD simulations or simply due to the distance between source and FWH surface.
For the latter case, this effect can be observed through a comparison of permeable surfaces with different sizes. In the limit of an infinite radius, the spherical surface would become a plane and the sources would be negligible on the surface elements far away from the true incident source (the monopole).
With that in mind, different spherical and incomplete planar FWH surfaces are analyzed for the present test case, as shown in Fig. \ref{fig:plane_monopole}(a). \red{The planar surfaces are square patches placed at $y=-0.5$ with side $L$ and outward normal vectors pointing along the negative $y$-axis.} 
Results of noise predictions for all surfaces are presented in Fig. \ref{fig:plane_monopole}(b). The solution obtained by the closed spherical surface with diameter $D=2$ is identical to the analytical one. In this case, all observers are outside of the permeable FWH surface. For the closed spheres with $D=4$ and $8$, some observers are positioned inside the FWH surface and, as expected, their numerical predictions are null, in agreement with the Heaviside function. For these spheres, the solutions computed at the observers placed outside the FWH surface match the analytical solution.

\begin{figure}[H]
	\centering
	\includegraphics[trim=2mm 2mm 2mm 2mm,clip,width=0.48\textwidth]{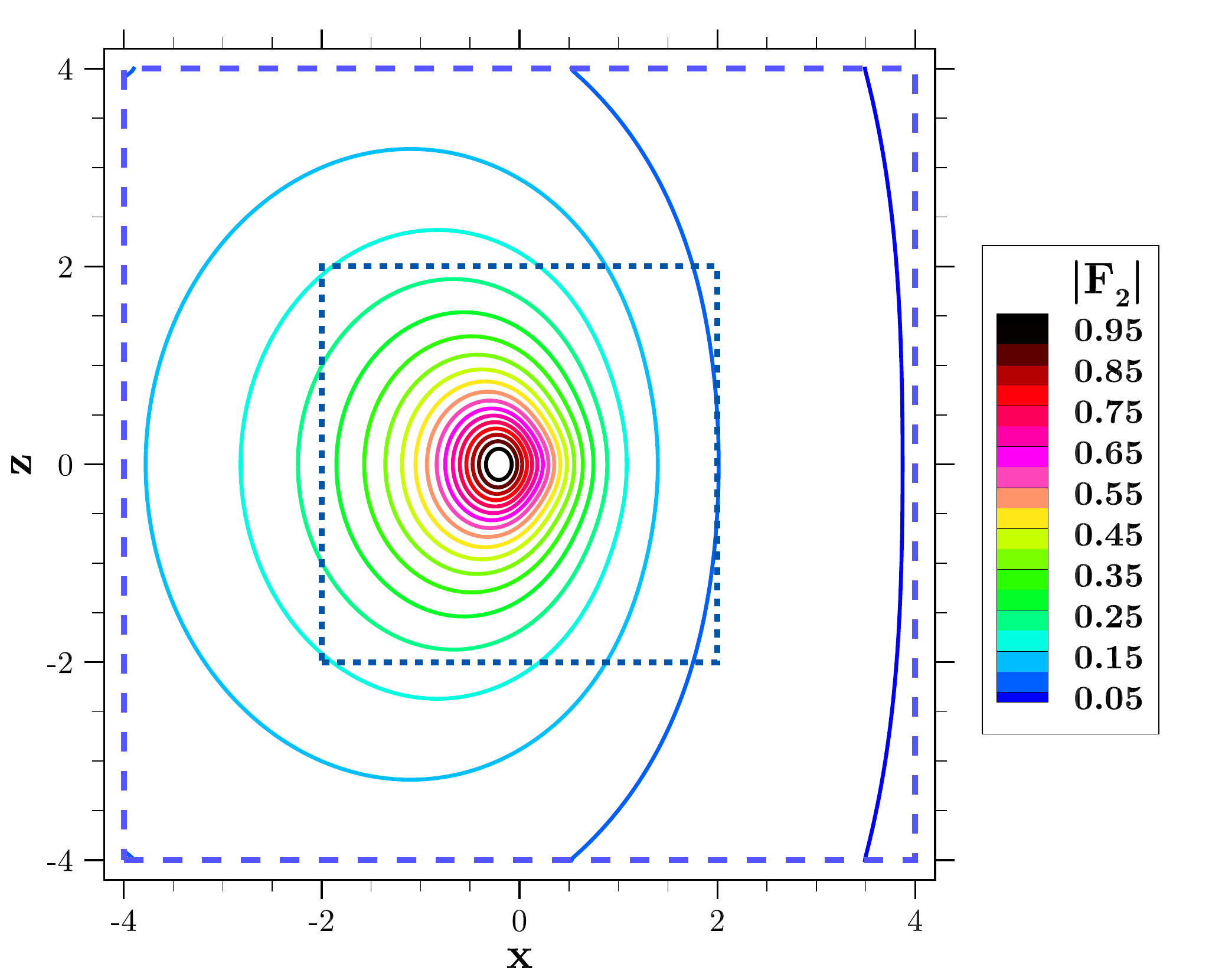}
	\caption{Distribution of normalized dipole source magnitude $\mathbf{|F_2(x,z)|}$ along the planar surfaces with $\mathbf{L=4}$ and $\mathbf{8}$, shown by the dotted and dashed lines, respectively.}
	\label{fig:plane_monopole_source}
\end{figure}

\red{Results for the planar surfaces show that, despite them being incomplete,} the noise predictions indeed converge to the analytical solution for observers at flyover positions (below $y=-0.5$, i.e., outside the FWH surface) when the larger surfaces with $L=8$ and $16$ are employed in the calculations. 
One important comment can be made about the deliberately small surface with $L=4$. At observer angles outside those accounted for by the integration surface, at $190 < \theta < 210$ deg. and $330 < \theta < 350$ deg, relevant acoustic sources are being neglected in the small permeable surface, which leads to large errors. Using the surface with $L=8$ already provides a good convergence towards the analytical solution.
This analysis shows that the prediction from an incomplete surface converges to the expected solution if, and only if, the source distribution on the surface decays significantly. \blue{In this sense, Fig. \ref{fig:plane_monopole_source} shows the surface distribution of the normalized $y$-component dipole source magnitude $|F_2(x,z)|$, as given by Eq. \ref{eq:dipole}, and it is the dominant equivalent source since both $n_1$ and $n_3$ are null for this planar setup}. The figure also presents the planar region $-4\leq x,z \leq 4$, which includes the patches with $L=4$ and $8$ depicted as dotted and dashed lines, respectively. The source distribution exhibits a decay towards the edges (dashed lines) of more than 85\% in magnitude. In summary, a closed FWH surface is required from a theoretical point of view but, in practice, an incomplete surface can be employed if wisely designed.


\subsection{Airfoil trailing edge noise due to vortex shedding}

The planar surface approach has an appeal for airframe noise predictions, where the sound radiation is particularly important at flyover and sideline observers. Furthermore, the plane option appears very convenient, as it can be designed not to intersect any part of the airplane or the support region of the quadrupoles. The plane (or any other similar shape) can be angled in order to be as close as possible to the noise source while remaining in the acoustic region and ensuring the line-of-sight requirement. One example of planar FWH surface application consists in the noise prediction from cavity flows \cite{Gloerfelt2003}. In this reference, good agreement between direct noise calculation from CFD and acoustic analogy is achieved using a planar FWH surface placed between the cavity and the observers.

Here, as an illustrative problem, we present results of noise prediction {from the two-dimensional flow over a NACA0012 airfoil at $3$ deg. angle of attack with a chord-based Reynolds number $Re = 10^4$, immersed in a uniform flow of $M = 0.2$. The flow simulation is performed using a $6^{th}$-order compact scheme for spatial discretization with an overset grid capability. In the present setup, an O-grid is generated along the airfoil to resolve the boundary layers while a Cartesian background grid solves the near acoustic field. No-slip adiabatic boundary conditions are enforced on the airfoil surface and characteristic plus sponge boundary conditions are applied in the farfield to minimize wave reflections. For the time marching, a hybrid approach is used where an implicit $2^{nd}$-order Beam \& Warming scheme is applied near the airfoil surface and an explicit $3^{rd}$-order Runge-Kutta scheme is used away from the solid wall. More details about the numerical methodology can be found in \cite{Bhaskaran2010,WolfJFM2012,Wolf2015}. 

Simulations are conducted using two different Cartesian background setups presented in Fig. \ref{fig:airfoil_mesh}. The grid shown in Fig. \ref{fig:airfoil_mesh}(a) has a more pronounced stretching towards the farfield boundaries, which is typical of airframe noise predictions of complex configurations (see \cite{Ricciardi2021}, for example). On the other hand, Fig. \ref{fig:airfoil_mesh}(b) presents a mesh with a more uniform resolution on the background block.
The Cartesian meshes employ $2.4$ and $6.5 \times 10^5$ grid points for the coarse and fine setups, respectively. Additionally, both setups have the same near-field O-mesh with $0.6 \times 10^5$ grid points. The time step for both simulations is $\Delta t = 0.001$ non-dimensionalized based on the sound speed and airfoil chord-length.
Using only the finer mesh, the DNC is computed all the way to the observers placed on a circle of radius $r_o = 5.0$, measured from the leading edge, as shown by the orange dotted line in Fig. \ref{fig:airfoil_mesh}(b). Considering the wavelength of the main tonal frequency, more than 20 points-per-wavelength are used to discretize the acoustic waves at the far-field. This is adequate to resolve both the main tone and its first harmonic given that high-order numerical schemes are employed.

Instantaneous solutions of the flowfield around the airfoil can be seen in Figs. \ref{fig:airfoil_solution}(a) and (b) for the coarser and finer grids, respectively. The gray contours present the acoustic field as the divergence of velocity, $|\nabla \cdot u| < 0.0025$, and the hydrodynamic field is presented with rainbow contours of $z$-vorticity, $|\omega_z| < 5$. As can be seen from the figures, the sound pattern makes the dominant role of the trailing edge obvious. Also, both wake solutions are similar since the near field meshes are designed with a good resolution downstream the airfoil. This is important to guarantee that the incident sources from vortex shedding for the two simulations will be similar. In both simulations, the vortices are suppressed near $x = 7$ (outside the circle of observers) by grid stretching and not by viscous diffusion. Regarding the acoustic field, the solution is compromised when using the coarser mesh due to its rapid stretching. For this reason, the FWH noise predictions will only be compared against the DNC computed using the finer mesh. 
Both simulations are performed for 250 convective time units based on the speed of sound after initial transients leave the domain and the signal becomes statistically stationary. Then, a Fourier transform is applied with a frequency resolution of $\Delta k = 0.01 \pi$ and no window function is employed in the transformation to avoid spectral leakage and further magnitude/energy corrections.
\begin{figure}[H]
	\centering
	\begin{subfigure}{0.48\textwidth}
		\includegraphics[width=0.99\textwidth]{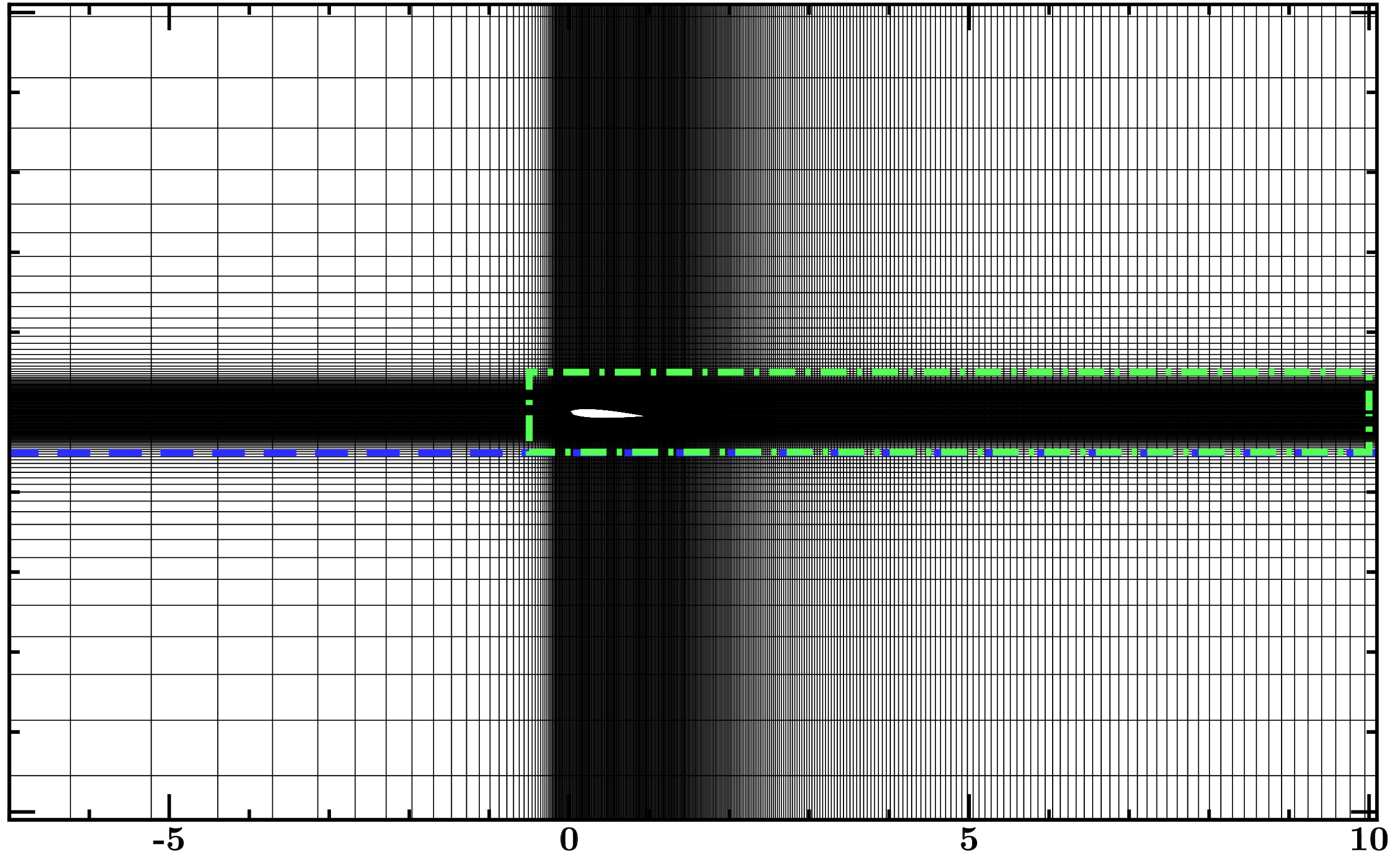}
		\subcaption{Coarser mesh}
	\end{subfigure}
	\begin{subfigure}{0.48\textwidth}
		\includegraphics[width=0.99\textwidth]{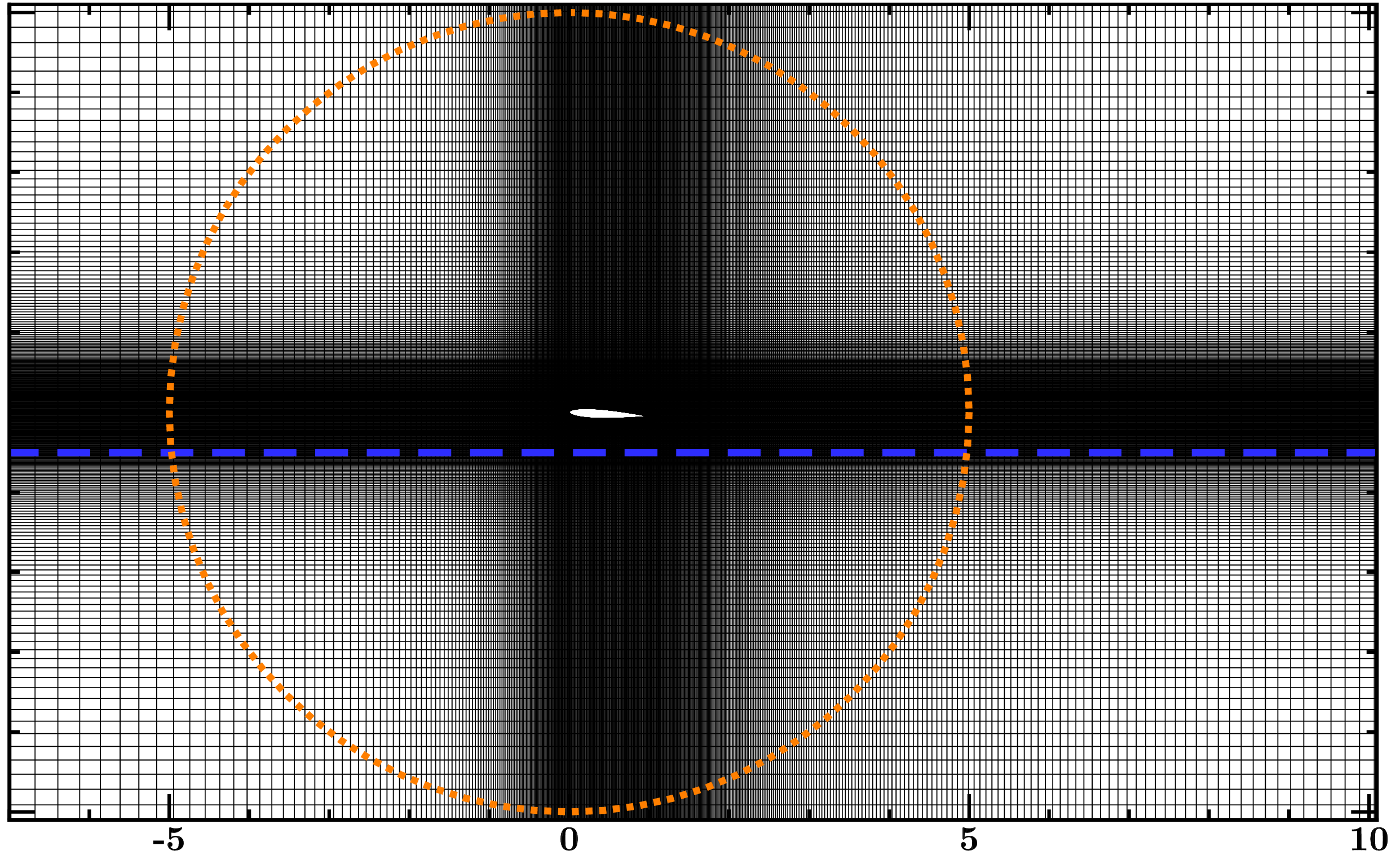}
		\subcaption{Finer mesh}
	\end{subfigure}
	\caption{Different Cartesian grid setups shown with a skip of 2 points in both the $\boldsymbol{x}$ and $\boldsymbol{y}$ directions.}
	\label{fig:airfoil_mesh}
\end{figure}
\begin{figure}[H]
	\centering
	\begin{subfigure}{0.48\textwidth}
		\includegraphics[width=0.99\textwidth]{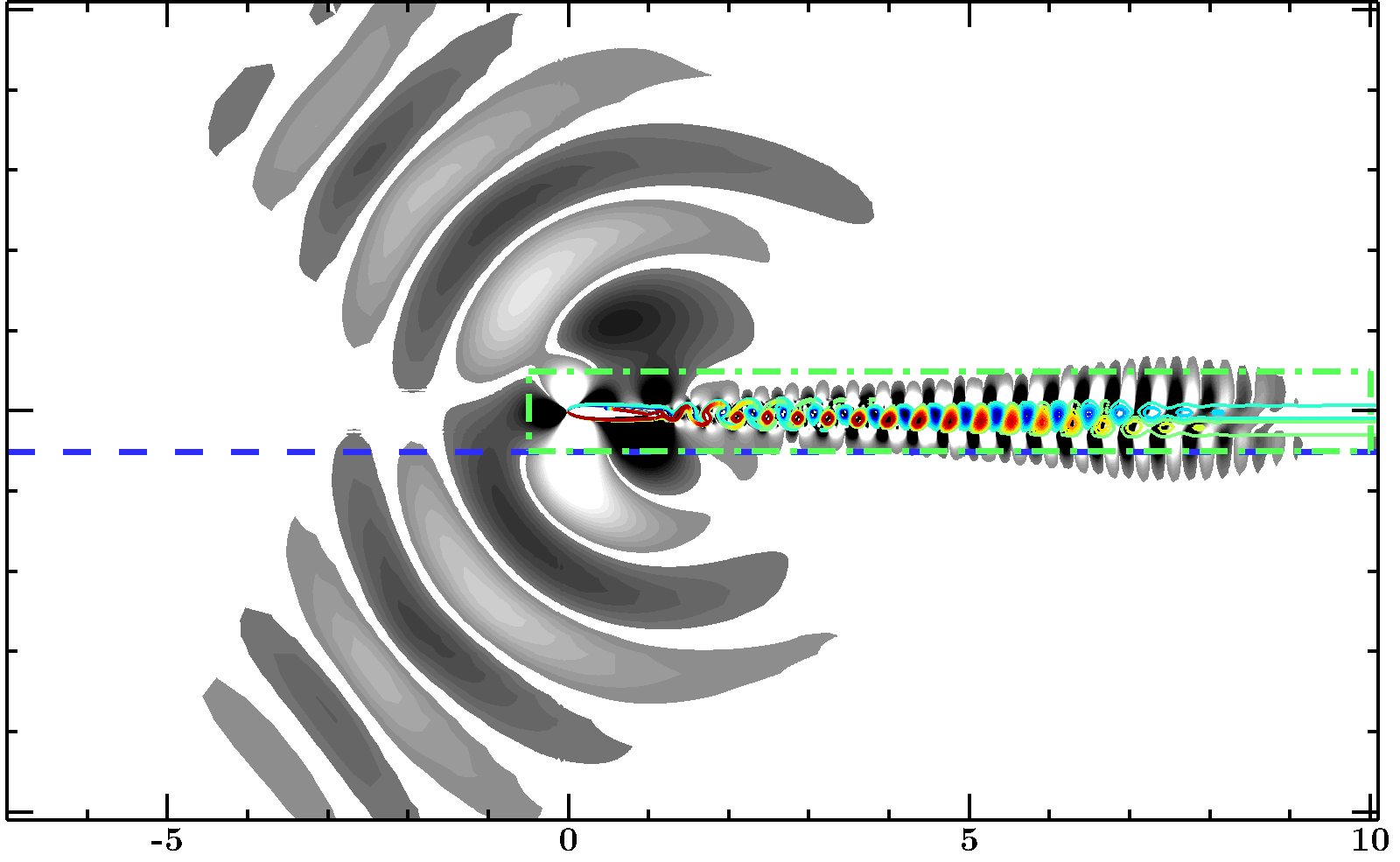}
		\subcaption{Coarser mesh}
	\end{subfigure}
	\begin{subfigure}{0.48\textwidth}
		\includegraphics[width=0.99\textwidth]{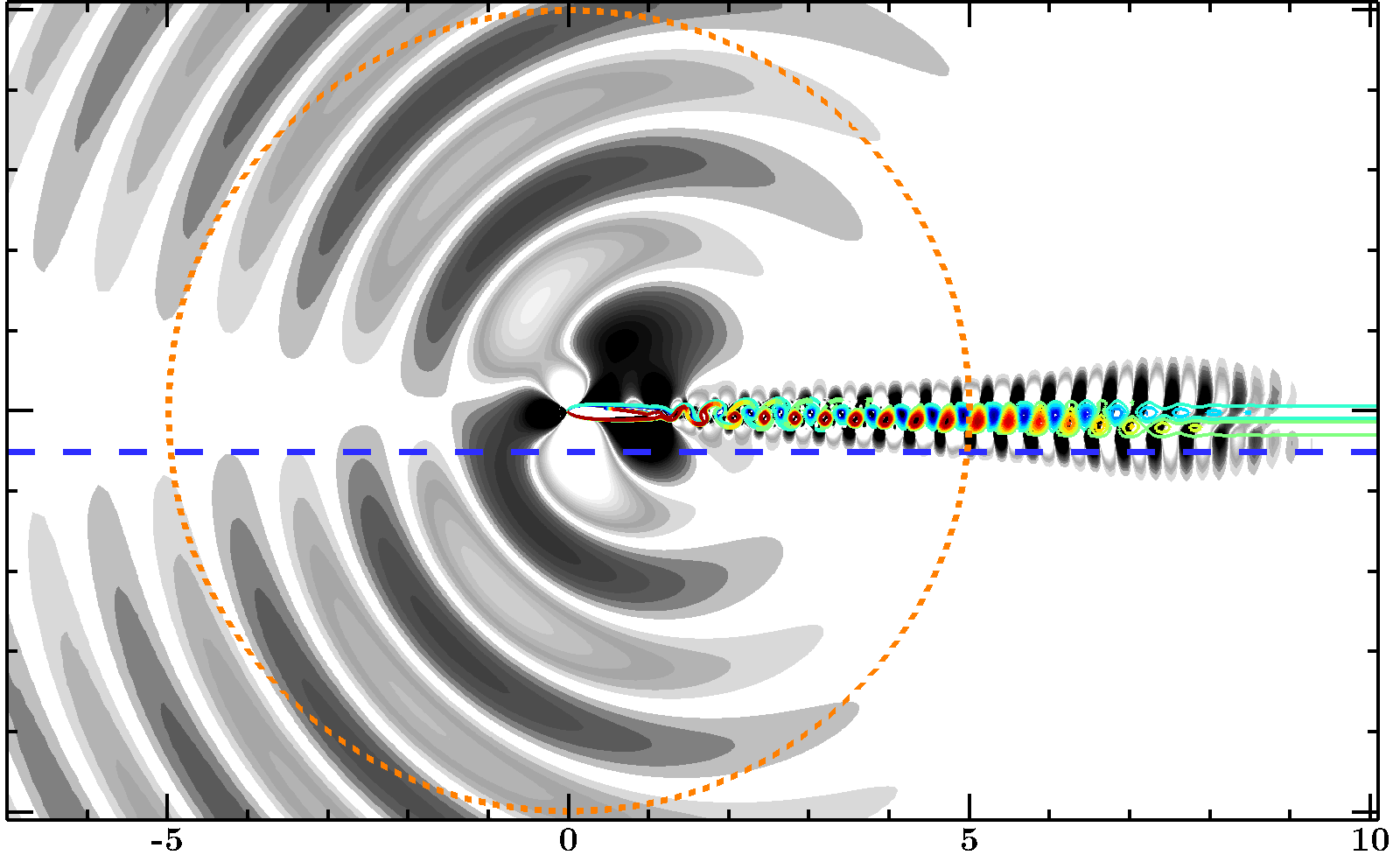}
		\subcaption{Finer mesh}
	\end{subfigure}
	\caption{Instantaneous solution of the flowfield around the NACA 0012 airfoil. The gray scale contours show the divergence of velocity ($\mathbf{|\nabla \cdot u| < 0.0025}$) and the rainbow contours depict the $\mathbf{z}$-vorticity ($\mathbf{|\omega_z| < 5}$).}
	\label{fig:airfoil_solution}
\end{figure}

\green{Several permeable FWH surfaces are tested and, for the coarser mesh, different approaches are employed including incomplete planar surfaces computed along the blue dashed line shown in the figure and a closed rectangular surface shown by the green dash-dotted line. In the latter, for sake of comparison, the influence of removing the downstream face is also analyzed. A planar surface is also tested with the finer mesh, in the same position as for the coarse grid.}
The acoustic prediction is performed for the tonal noise peak at $k \approx 3.08$. In the initial assessment, only the solid surface and rectangular permeable surfaces of the coarser mesh, with and without the downstream closing face, are considered. Starting with the integration surface extending 1 chord downstream of the airfoil, results in Fig. \ref{fig:airfoil_closed_permeable}(a) show a mismatch between closed permeable FWH surface and DNC. This occurs because the back face is too close to the airfoil and leads to the contamination problems discussed in the literature. It should be clarified that the solution is obtained with a single endcap. The solid surface slightly underpredicts the noise, which indicates that quadrupole sources are small but non-negligible. 
The best result is achieved with the open rectangular surface, i.e., without the back permeable face, in agreement with the DNC and apparently without issues with hydrodynamic contamination.
Extending the permeable FWH surface further downstream, the noise prediction matches the DNC solution when the back face is placed 9 chords downstream of the airfoil, as presented in Fig. \ref{fig:airfoil_closed_permeable}(b). This is a consequence of damping the hydrodynamic fluctuations along the wake, as illustrated in Fig. \ref{fig:airfoil_solution}(a). The numerical dissipation of flow structures cancels an error here, but cannot be viewed as a reliable solution in general.
\begin{figure}[H]
	\centering
	\begin{subfigure}{0.45\textwidth}
		\includegraphics[width=0.90\textwidth]{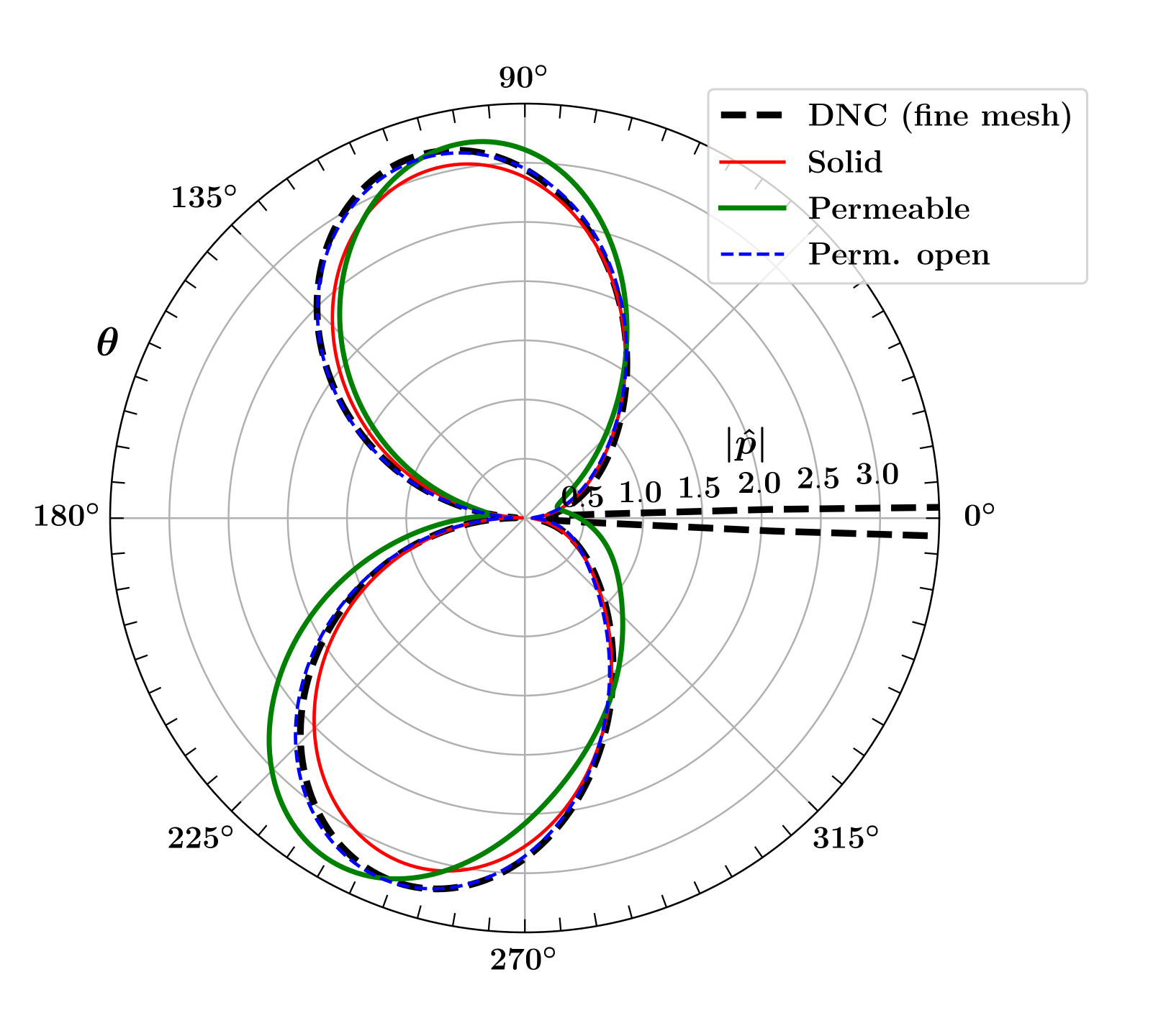}
		\subcaption{Back face at x = 2.}
	\end{subfigure}
	\begin{subfigure}{0.45\textwidth}
		\includegraphics[width=0.90\textwidth]{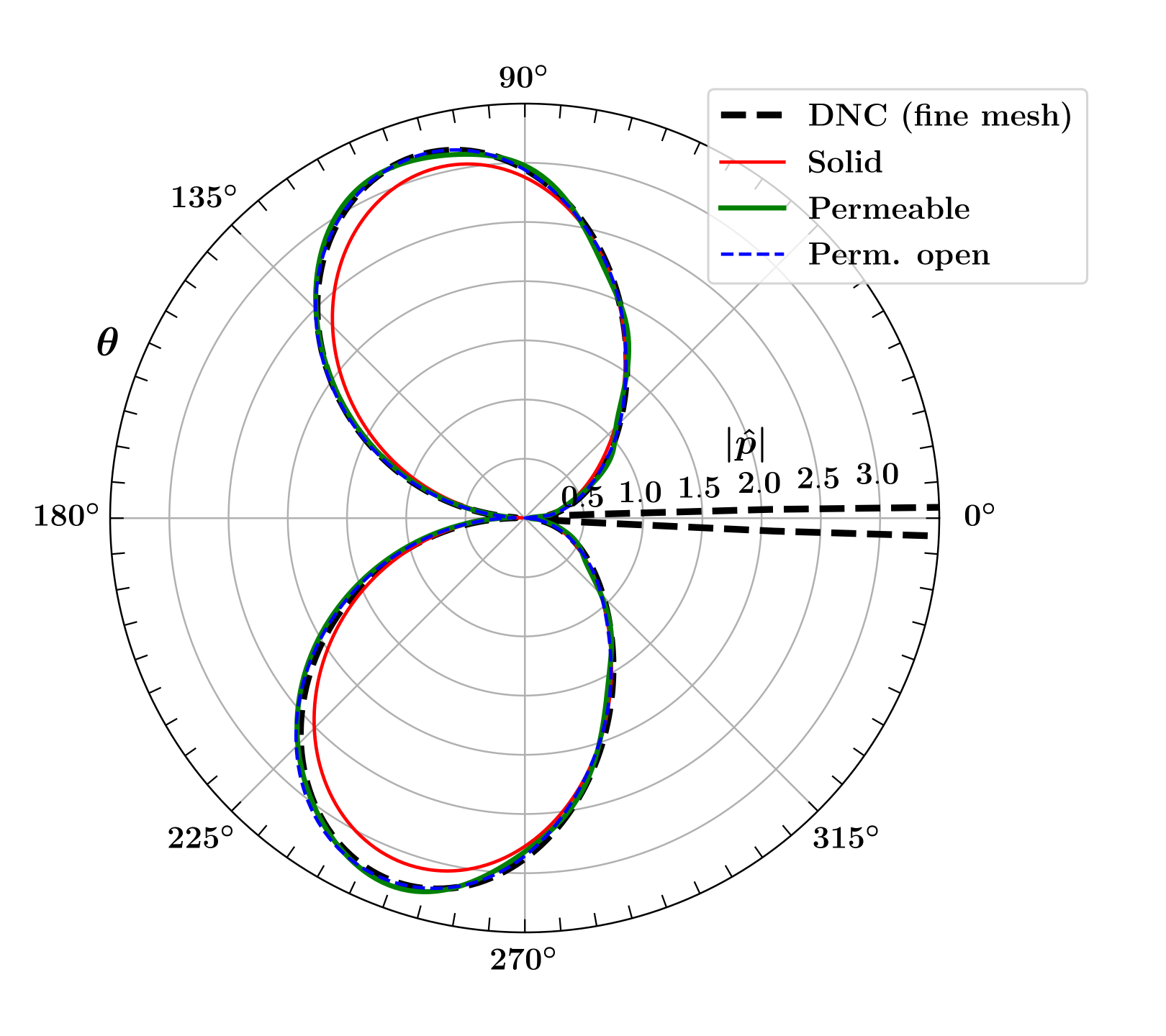}
		\subcaption{Back face at x = 10.}
	\end{subfigure}
	\caption{Directivity plots of solid and permeable FWH surfaces.}
	\label{fig:airfoil_closed_permeable}
\end{figure}

Next, the finite planar FWH surfaces are tested to compute the noise radiation to flyover observers. Several configurations are analyzed where, in the first case, the plane only covers the airfoil unitary chord, extending from $0 < x < 1$. In the second setup, the surfaces extend from one chord upstream the airfoil to different downstream distances, which are varied in order to assess convergence of results. In a third setup, the open surfaces extend one chord downstream the airfoil and the influence of upstream length is investigated. Results for the second setup are shown in Figs. \ref{fig:airfoil_plane_linear}(a) and \ref{fig:airfoil_plane_log}(a), while solutions obtained by the third setup are presented by Figs. \ref{fig:airfoil_plane_linear}(b) and \ref{fig:airfoil_plane_log}(b). As the planes are extended away from the airfoil, they eventually reach the observer locations. In these figures, the solutions presented along the gray shaded regions should not be considered since they are computed for observers placed above the planar surfaces.

If the plane only covers the exact extension of the airfoil, the noise prediction is significantly off compared to the DNC. Then, varying the plane size downstream exhibits a mismatch which is independent of the surface size, as presented in Fig. \ref{fig:airfoil_plane_linear}(a). Still, the levels are close to the expected values. Furthermore, as the plane extends along the aerodynamic wake, hydrodynamic fluctuations contaminate the surface and lead to a more oscillatory pattern which can be seen for $-1<x<6$ in the magenta dashed line. This occurs because the planar surface is truncated inside the source hydrodynamic region. If the plane is extended beyond the wake ($-1<x<10$) the truncation is avoided and the ripples vanish providing a solution with good agreement when compared to that for the plane positioned at $-1<x<2$.
Figure \ref{fig:airfoil_plane_linear}(b) shows that the results from acoustic analogy converge to the expected solution as more of the upstream traveling acoustic waves are properly captured when the planar surface is extended upstream. This indicates that FWH acoustic analogy can still be used with an incomplete surface.

In noise predictions, results are typically presented in terms of the power spectral density, PSD, measured in logarithmic scale. In this sense, Fig. \ref{fig:airfoil_plane_log} shows that the deviations between DNC and the solutions computed by the incomplete surfaces at $-1<x<2$ are smaller than 2dB when measured along the arc $225 < \theta < 315$ deg. These angles represent flyover observers when the planar surfaces extend at least 1 chord on both directions from the airfoil. Solutions are even more accurate if larger upstream planes are employed, even when they are placed on regions with pronounced grid stretching. Hence, we expect that the present planar FWH approach can be useful for airframe noise predictions of complex configurations at flyover and sideline observer positions.
\begin{figure}[H]
	\centering
	\begin{subfigure}{0.49\textwidth}
		\includegraphics[trim=2mm 18mm 1mm 18mm,clip,width=0.90\textwidth]{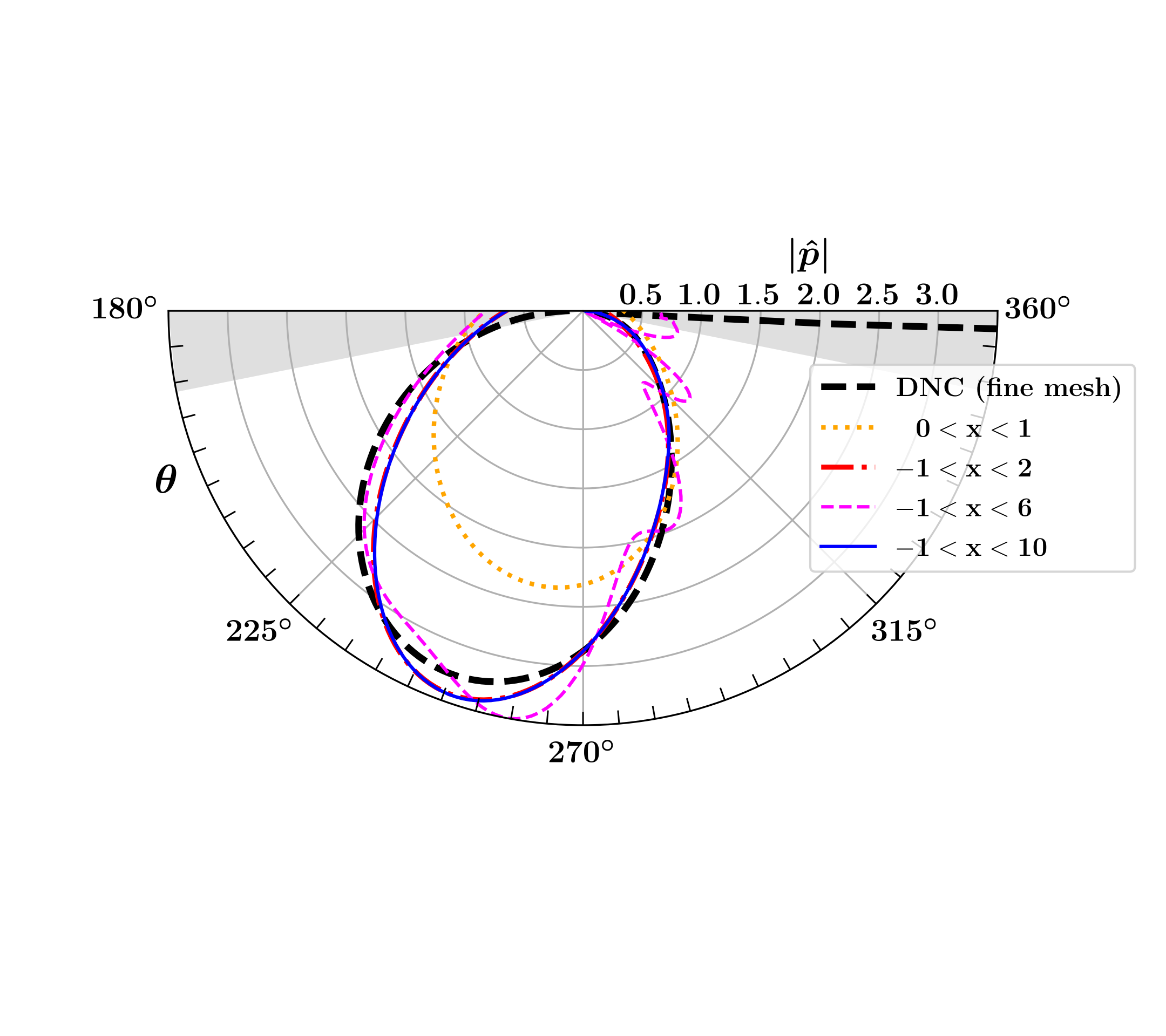}
		\subcaption{Planes extending along the downstream direction.}
	\end{subfigure}
	\begin{subfigure}{0.49\textwidth}
		\includegraphics[trim=2mm 18mm 1mm 18mm,clip,width=0.90\textwidth]{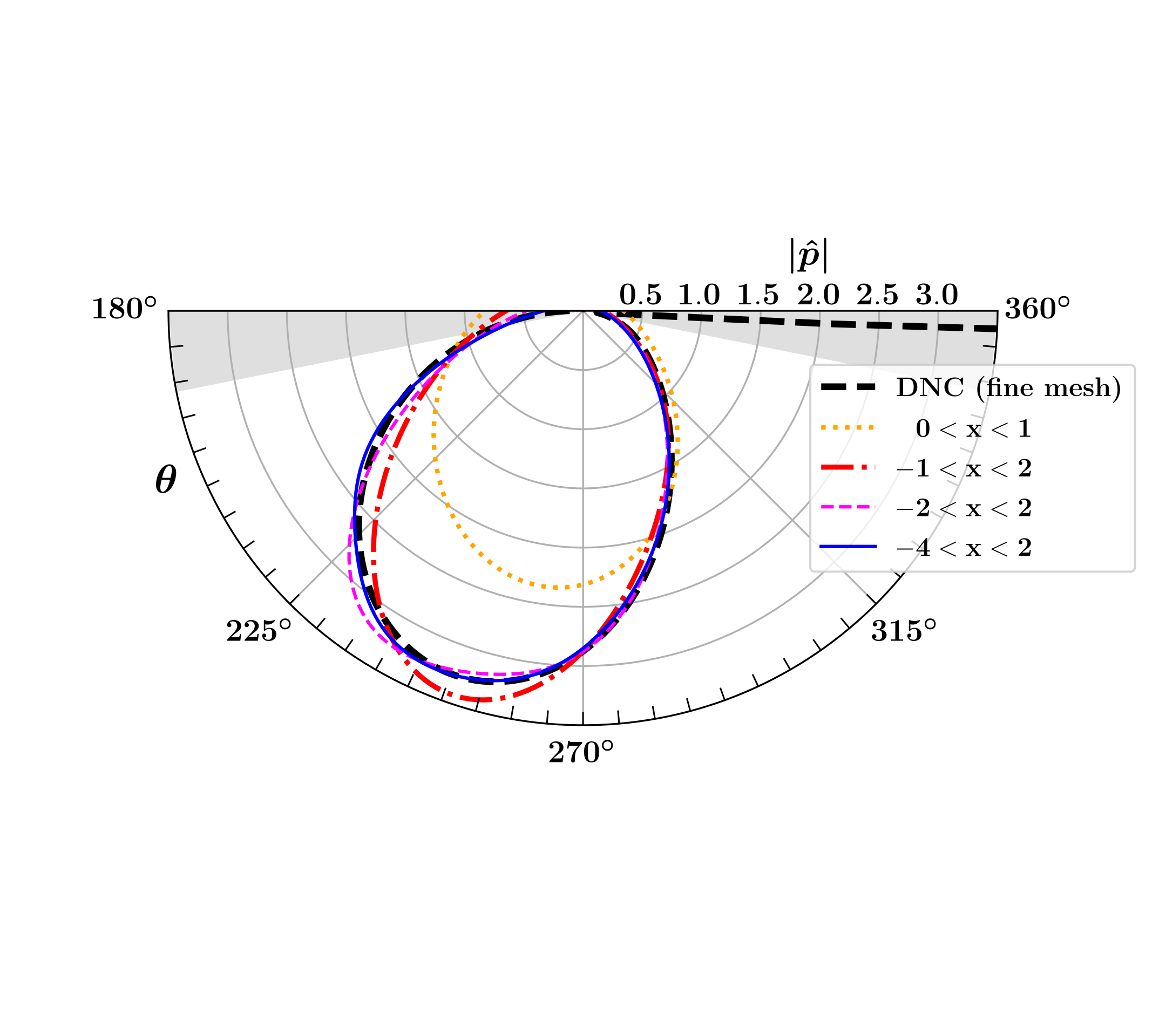}
		\subcaption{Planes extending along the upstream direction.}
	\end{subfigure}
	\caption{Directivities for incomplete permeable FWH surfaces.}
	\label{fig:airfoil_plane_linear}
\end{figure}
\begin{figure}[H]
	\centering
	\begin{subfigure}{0.49\textwidth}
		\includegraphics[width=0.90\textwidth]{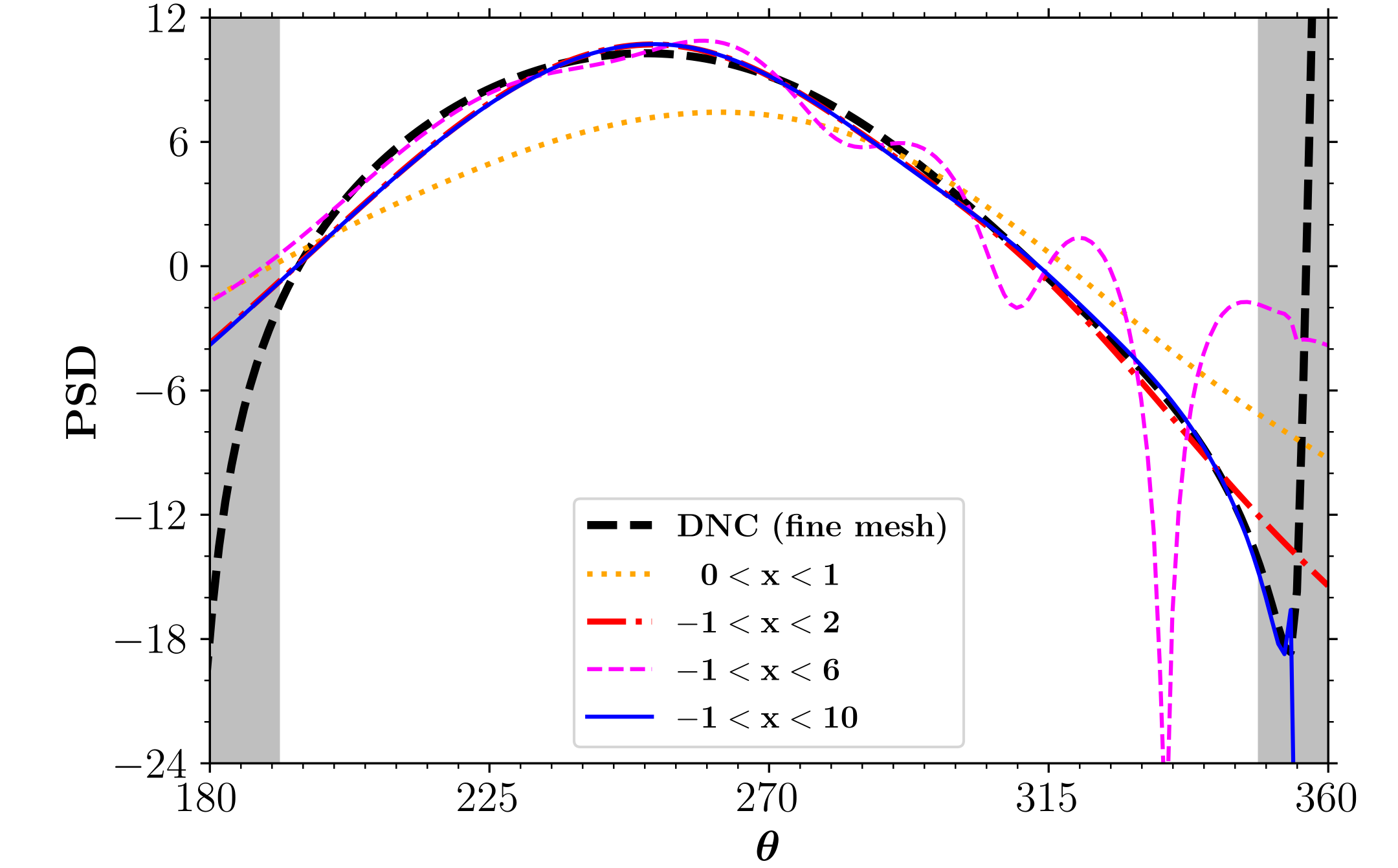}
		\subcaption{Plane extending along the downstream direction.}
	\end{subfigure}
	\begin{subfigure}{0.49\textwidth}
		\includegraphics[width=0.90\textwidth]{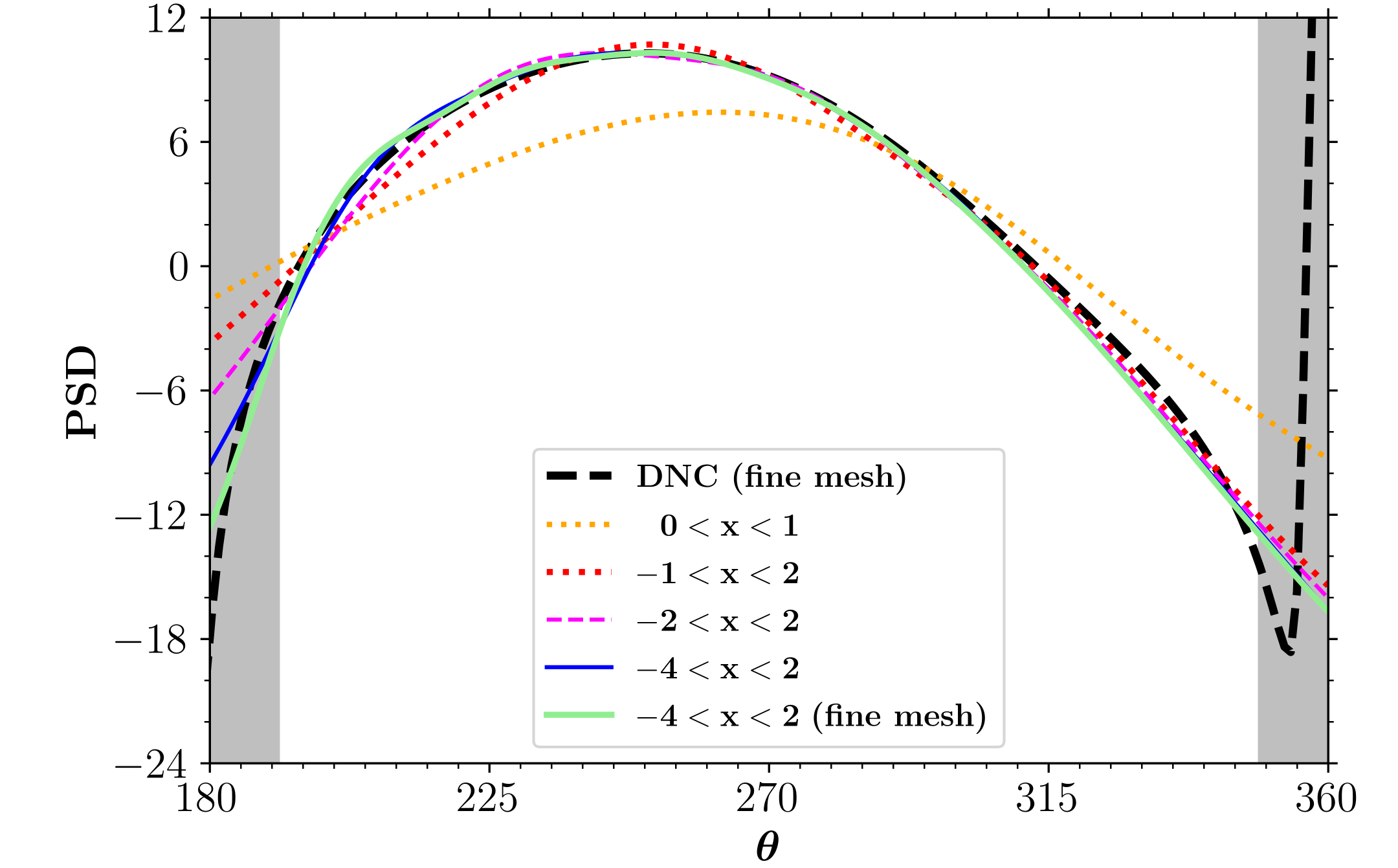}
		\subcaption{Plane extending along the upstream direction.}
	\end{subfigure}
	\caption{Directivities computed as PSD for incomplete permeable FWH surfaces (log scale).}
	\label{fig:airfoil_plane_log}
\end{figure}

Time signals at the far-field are computed using an inverse Fourier transform of the complex-valued noise prediction using the FWH acoustic analogy and compared against the reference value from the DNC. Since the latter is computed using the finer mesh and the former employs the coarse mesh, a phase correction is applied to the DNC for a proper comparison. Results for the flyover observer at $\theta = 270$ deg. are presented in Fig. \ref{fig:airfoil_plane_time}(a). There, it is possible to see a good agreement between the noise prediction using the different FWH surfaces and the DNC, with the exception of the minimal surface ranging from $0 < x < 1$. The spectra in Fig. \ref{fig:airfoil_plane_time}(b) depict the whole bandwidth of frequencies considered in the noise propagation and a good agreement is observed for the main tone and its first harmonic. For the higher harmonics, numerical dissipation reduces the pressure level of the DNC compared to the FWH predictions. Deviations are observed on the broadband levels for the lower wavenumbers due to errors in the Fourier analysis, but it is important to note that these errors are four orders of magnitude lower compared to the main tonal peak.

\begin{figure}[H]
	\centering
	\begin{subfigure}{0.51\textwidth}
		\includegraphics[trim=0mm 0mm 0mm 0mm,clip,width=0.80\textwidth]{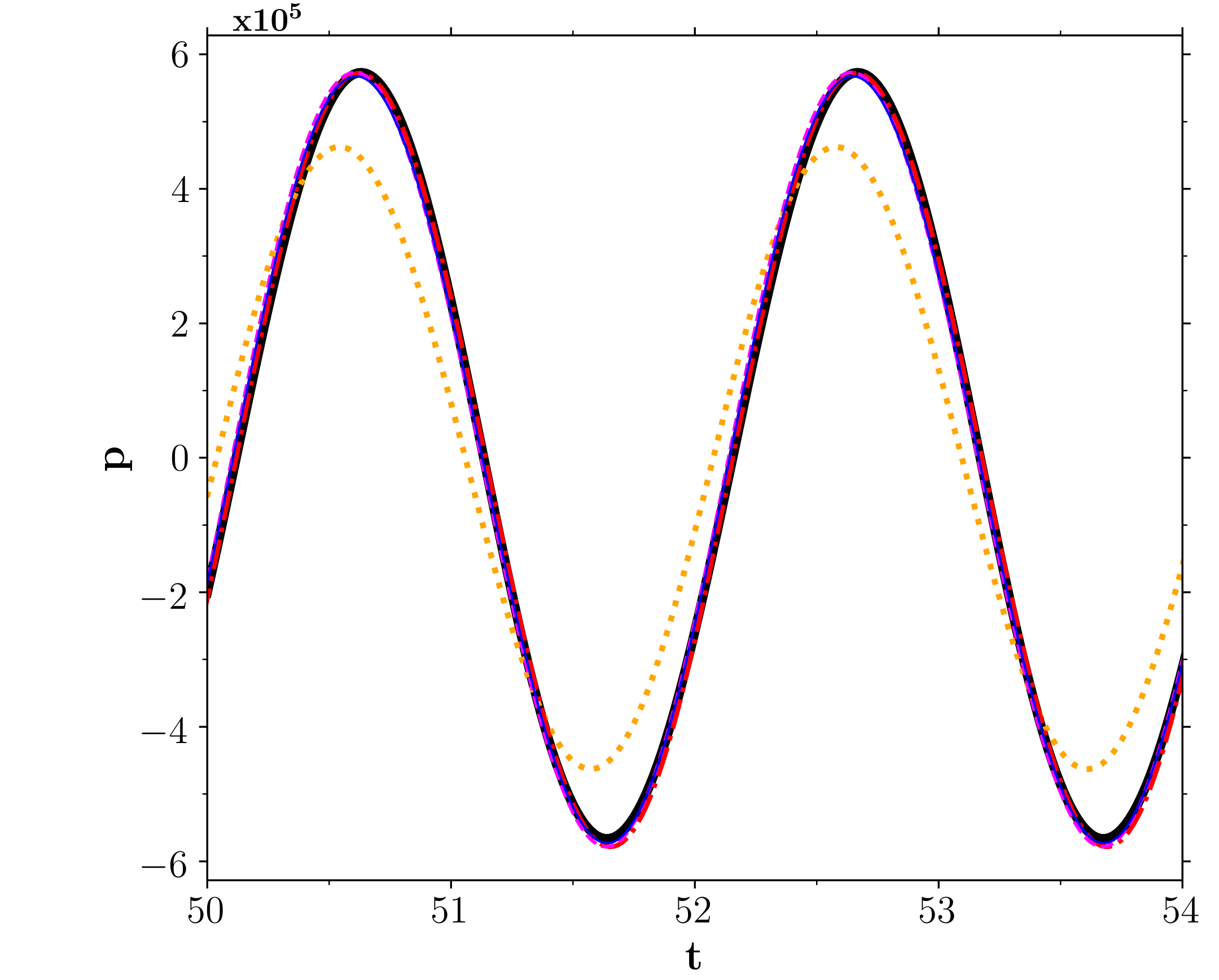}
		\subcaption{Time signals obtained from inverse Fourier transform.}
	\end{subfigure}
	\begin{subfigure}{0.48\textwidth}
		\includegraphics[trim=0mm 0mm 0mm 0mm,clip,width=0.85\textwidth]{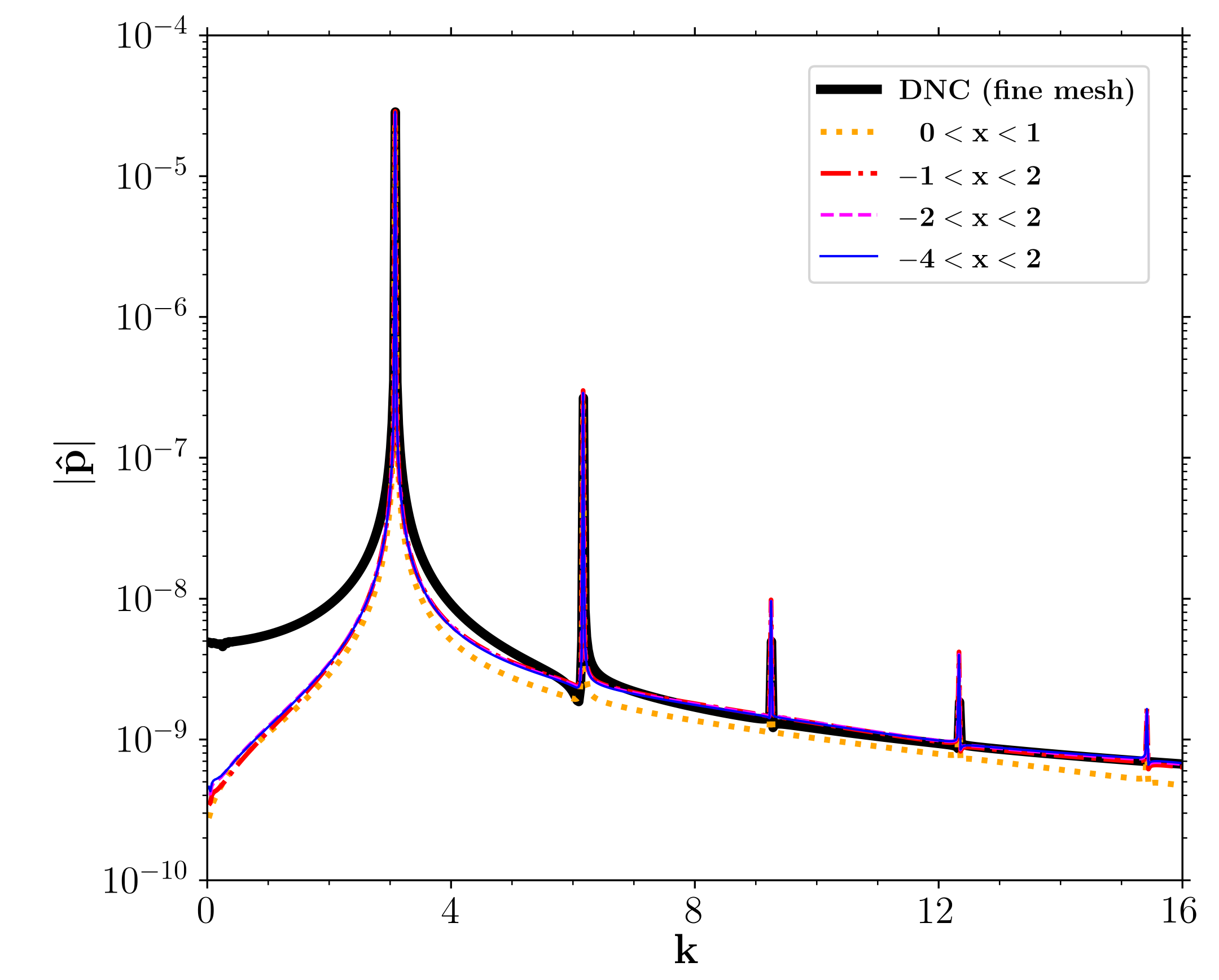}
		\subcaption{Frequency spectra of acoustic pressure.}
	\end{subfigure}
	\caption{Temporal signals and their spectral content obtained from the DNC and incomplete FWH surfaces for observers at $\boldsymbol{\theta = 270}$ deg.}
	\label{fig:airfoil_plane_time}
\end{figure}

\section{Conclusions}
\label{discussion}

This work is motivated by CFD simulations from a realistic landing gear performed modeling only the half bottom of the aircraft fuselage \cite{Ricciardi2021}. Hence, in this previous analysis, the aeroacoustic predictions had to employ incomplete FWH surfaces. In the present study, a discussion on the closed surface requirement is provided for the frequency domain FWH formulation, although we believe that the approach presented here can also be applied for the time-domain methodology. We show that all sources computed on the closed permeable surface may play an important role if the acoustic prediction is sought at observer positions distributed along the entire circular arc. However, results obtained for 2D and 3D model problems show that, if wisely designed, a permeable FWH surface composed only by a finite patch, i.e., an incomplete surface, can be employed to accurately predict the noise at specific observer positions. In this case, the surface must be placed between the line of sight from the sources to the observers, for example, in flyover or sideline locations. In the current applications, results from finite surfaces agree with the expected values as long as the source magnitudes decay on elements distant from the true incident source, i.e., the airframe. Hence, in order for this approach to be valid, the sources cannot be truncated. The planar setup is beneficial in airframe noise applications since it avoids contamination by quadrupole sources crossing the boundaries and also may simplify the surface design.

\section*{Acknowledgments}
	The authors of this work would like to acknowledge Fun\-da\-\c{c}\~{a}o de Amparo \`{a} Pesquisa do Estado de S\~{a}o Paulo, FAPESP, for supporting the present work under research grants No.\ 2013/08293-7 and 2018/11835-0. The authors thank LNCC-Cluster SDumont (Project SimTurb) for providing the computational resources used in this study.
	
\bibliography{biblio}

\end{document}